\documentclass{INTERSPEECH2023}

\usepackage{amsfonts}
\usepackage{multirow}
\usepackage{bbding}
\usepackage{cite}
\usepackage{subfigure}
\usepackage{mathrsfs}

\interspeechcameraready

\def\scaleencmodule{ScaleFuser}
\def\scaledecmodule{ScaleInterMG}
\def\fusmodule{ConSM}
\def\model{MC-SpEx}

\title{\model{}: Towards Effective Speaker Extraction with Multi-Scale Interfusion and Conditional Speaker Modulation}
\name{Jun Chen$^{1,2,\dagger}$\thanks{$^{\dagger}$ Work conducted when the first author was intern at Tencent.}, Wei Rao$^{2}$, Zilin Wang$^1$, Jiuxin Lin$^1$, Yukai Ju$^2$, Shulin He$^2$ \\ \textit{Yannan Wang$^2$, Zhiyong Wu$^{1,3,*}$\thanks{$^{*}$ Corresponding author.}}}
\address{
  $^1$Shenzhen International Graduate School, Tsinghua University, Shenzhen, China\\
  $^2$Tencent Ethereal Audio Lab, Tencent, Shenzhen, China \\
  $^3$The Chinese University of Hong Kong, Hong Kong SAR, China
  }
\email{y-chen21@mails.tsinghua.edu.cn, ellenwrao@tencent.com, zywu@sz.tsinghua.edu.cn}

\begin{document}

\maketitle

\begin{abstract}
\vspace{-0.1cm}
The previous SpEx+ has yielded outstanding performance in speaker extraction and attracted much attention.
However, it still encounters inadequate utilization of multi-scale information and speaker embedding.
To this end, this paper proposes a new effective speaker extraction system with multi-scale interfusion and conditional speaker modulation (\fusmodule{}), which is called \model{}.
First of all, we design the weight-share multi-scale fusers (\scaleencmodule{}s) for efficiently leveraging multi-scale information as well as ensuring consistency of the model's feature space.
Then, to consider different scale information while generating masks, the multi-scale interactive mask generator (\scaledecmodule{}) is presented.
Moreover, we introduce \fusmodule{} module to fully exploit speaker embedding in the speech extractor.
Experimental results on the Libri2Mix dataset demonstrate the effectiveness of our improvements and the state-of-the-art performance of our proposed \model{}.
\end{abstract}

\vspace{-0.2cm}
\section{Introduction}
\vspace{-0.1cm}
Speech separation, commonly known as the cocktail-party problem, is a fundamental challenge in the field of speech processing that intends to separate each source signal from the mixed speech of multiple speakers.
Most studies on speech separation are limited by the requirement for prior knowledge of the number of speakers, and additionally they entail addressing the global permutation ambiguity challenge \cite{xu2020spex} to channel the correct speaker to the correct output voice stream.
In order to avoid these constraints, speaker extraction is proposed as a strategy that only extracts target speaker's speech from the mixture according to the reference speech from target speaker.
It can be used in a variety of downstream applications, including automatic speech recognition (ASR), real-time communication (RTC) and speaker diarization, just to name a few.

Motivated by humans' top-down auditory attention to the target speaker
\cite{mesgarani2012selective, kaya2017modelling}, deep learning-based speaker extraction methods \cite{zmolikova2017speaker, xu2019time, vzmolikova2019speakerbeam, delcroix2019compact, wang2020voicefilter, delcroix2020improving, he2020speakerfilter, ju2022tea, ju2023tea} primarily adopt a two-subnet architecture consisting of a speaker encoder and a speech extractor, 
in which the speaker encoder models the speaker representation of the target speaker, and then directs the speech extractor to extract the speech signal belonging to the target speaker.
While following the previous practice, SpEx+ \cite{ge2020spex+} further introduces a twin speech encoder with shared parameters to capture multi-scale speech features directly from the speech waveform for the speaker encoder and speech extractor, and reverts their processed multi-scale features to the waveform via the speech decoder.
By doing so, a uniform latent feature space containing multi-scale information is introduced for both the subnets of speaker encoder and the speech extractor, resulting in outstanding results for SpEx+.

Despite the impressive performance, SpEx+ is still not perfect.
The remarkable achievements of Conformer in source separation \cite{li2022ead, chen2021continuous} and speaker verification \cite{zhang2022mfa} demonstrate that the effective fusion of multi-scale speech information enables neural networks to leverage more comprehensive acoustic features, which is conducive to the model performance.
Nevertheless, in SpEx+, the fusion of the multi-scale information extracted by speech encoder could be further boosted.
And, while the twin speech encoder of SpEx+ considers the consistency of the two subnets' feature space \cite{ge2020spex+}, the original independent fusion modules within each subnet do not take this into account so far.
Besides, when decoding the mask, SpEx+ merely employs three separate branches to generate three masks for different scales, which fails to adequately combine and utilize information in multiple scales.
Furthermore, it has been pointed out that the speaker embedding, which is concatenated with the frame-level speech features, cannot be sufficiently exploited by the stacked temporal convolutional network (TCN) blocks \cite{wang2021neural}.
We argue that this underutilization of speaker embedding arisen in SpEx+ may compromise the model's discrimination for target speaker in mixed speech, consequently leading to a limited capability of the speaker extraction system.

In this paper, to tackle the insufficient utilization of multi-scale information as well as speaker embedding in SpEx+, we propose an efficient speaker extraction system called \model{}.
We design the \scaleencmodule{} to more effectively leverage multi-scale information extracted from the speech waveform.
Afterwards, for the consistency of two subnets' feature space, we incorporate the newly designed \scaleencmodule{}s with shared parameters into the twin speech encoders to obtain twin multi-scale fusion speech encoders.
Correspondingly, the \scaledecmodule{} is also presented to substitute the original three independent branches, so as to take the different scale information into account when generating masks.
Together with the original speech decoder, the \scaledecmodule{} constitutes the multi-scale interactive speech decoder.
Furthermore, we introduce \fusmodule{} module to fully blend speaker embedding into the speech extractor.
Experimental results show that \model{} significantly outperforms the performance of our baseline SpEx+ in extracting the speech signals of the target speaker, which confirms the effectiveness of our improvements. 
Moreover, our proposed \model{} also yields state-of-the-art results for the speaker extraction task on the Libri2Mix dataset \cite{cosentino2020librimix}.

\vspace{-0.1cm}
\begin{figure}[!htbp]
	\centering
    \includegraphics[width=0.97\linewidth]{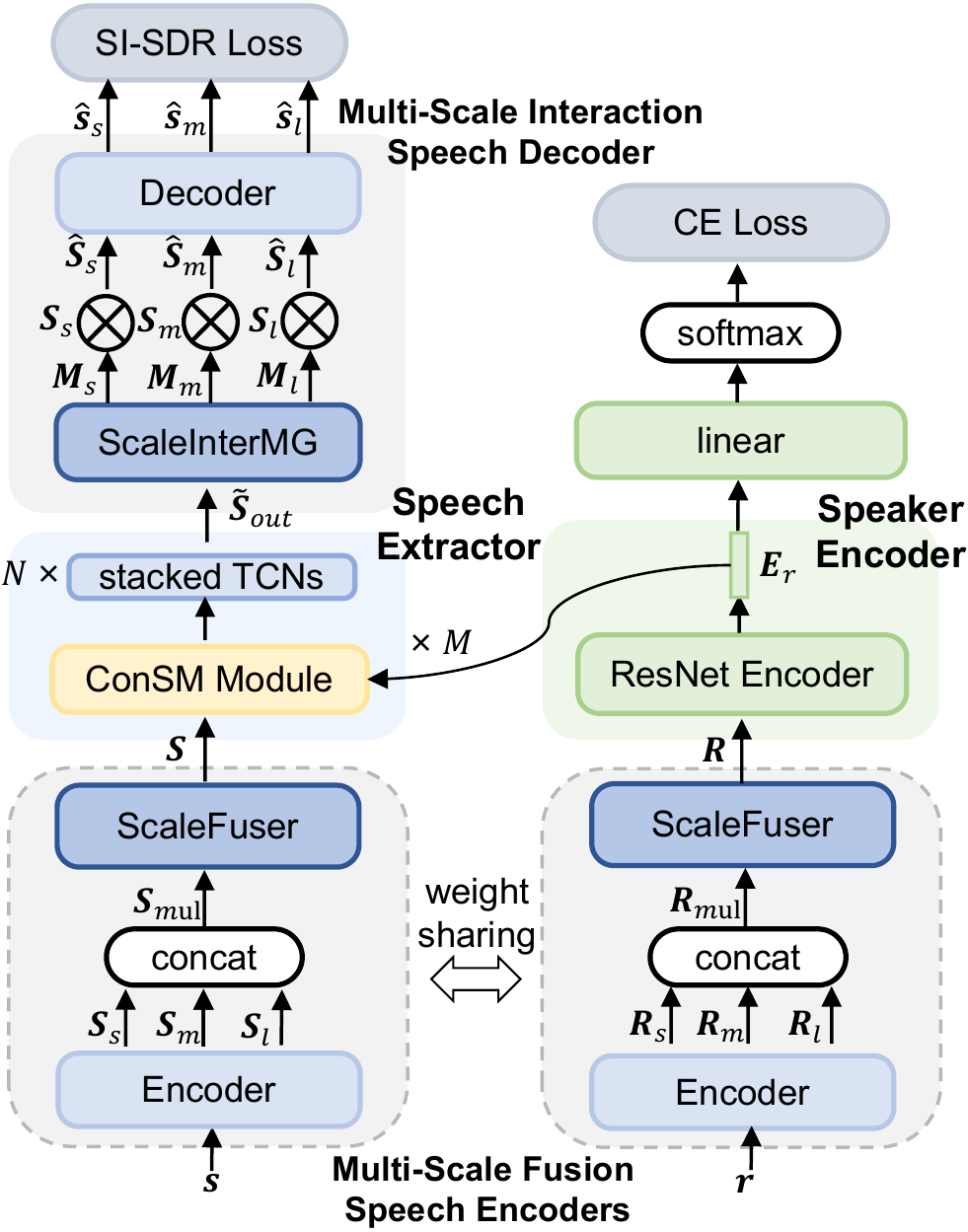}
    \vspace{-0.2cm}
	\caption{The overall diagram of the \model{}.
	The dotted border represents the module with shared weights.
	The ``$\otimes$" means element-wise multiplication.
	The ``SI-SDR Loss" and ``CE Loss" refer to the scale-invariant signal-to-distortion ratio loss and cross-entropy loss in multi-task learning \cite{xu2020spex}.
	}

	\label{fig:totalarch}
	\vspace{-0.4cm}
\end{figure}

\section{Methodology}
\vspace{-0.1cm}
As shown in Figure \ref{fig:totalarch}, \model{} is composed of four main parts: multi-scale fusion speech encoders, speaker encoder, speech extractor, and multi-scale interaction speech decoder.
One of the subnets, the speaker encoder, uses a residual network (ResNet) encoder \cite{zeinali2019but} comprising of stacked ResNet blocks and a pooling layer.
As the other subnet, the speech extractor consists of $M$ groups of speaker-guided stacked TCN blocks.
Inside each group, there are the \fusmodule{} module in the front end and $N$ stacked TCNs with exponentially growing dilation factors $\{2^n\}(n \in \{0, \cdots, N-1\})$.

The proposed model takes the mixed speech waveform $\mathbf{s}$ and reference speech waveform $\mathbf{r}$ as inputs.
In the weight-share multi-scale fusion speech encoders, the encoder respectively extracts multi-scale speech features $\mathbf{S}_{mul} = [\mathbf{S}_{s}, \mathbf{S}_{m}, \mathbf{S}_{l}]$ and $\mathbf{R}_{mul} = [\mathbf{R}_{s}, \mathbf{R}_{m}, \mathbf{R}_{l}]$ from waveforms $\mathbf{s}$ and $\mathbf{r}$, 
where ``$[,]$" denotes the concatenation operation and the subscripts $s$, $m$ and $l$ refer to small, middle, and large scales respectively as in \cite{ge2020spex+}.
And then, the \scaleencmodule{}s are responsible for fusing the multi-scale information, in which $\mathbf{S}_{mul}$ and $\mathbf{R}_{mul}$ are fused into $\mathbf{S}$ and $\mathbf{R}$, respectively.
$\mathbf{R}$ is fed to speaker encoder and then we get the speaker embedding $\mathbf{E}_r$ from it to represent the characteristics of the target speaker.
The speech extractor, under the direction of $\mathbf{E}_r$, outputs processed speech features $\widetilde{\mathbf{S}}_{out}$ based on another input $\mathbf{S}$.
Eventually, in multi-scale interaction speech decoder, the \scaledecmodule{} interactively generates multi-scale receptive masks $\mathbf{M}_{s}$, $\mathbf{M}_{m}$ and $\mathbf{M}_{l}$ from $\widetilde{\mathbf{S}}_{out}$.
After the element-wise multiplication between multi-scale masks and the corresponding multi-scale features in $\mathbf{S}_{mul}$, we obtain the speech features $\widehat{\mathbf{S}}_{s}$, $\widehat{\mathbf{S}}_{m}$ and $\widehat{\mathbf{S}}_{l}$ predicted by the model, and these features are transformed by decoder to gain the estimated speech waveform $\hat{\mathbf{s}}_{s}$, $\hat{\mathbf{s}}_{m}$ and $\hat{\mathbf{s}}_{l}$.
These predicted speech waveforms and $\mathbf{E}_r$ are eventually used to calculate the multi-task learning loss for training the model as described in \cite{xu2020spex}.
We will elaborate on our improvements in the following subsections.

\begin{figure}[tp]
    \centering
    \includegraphics[width=1.0\linewidth]{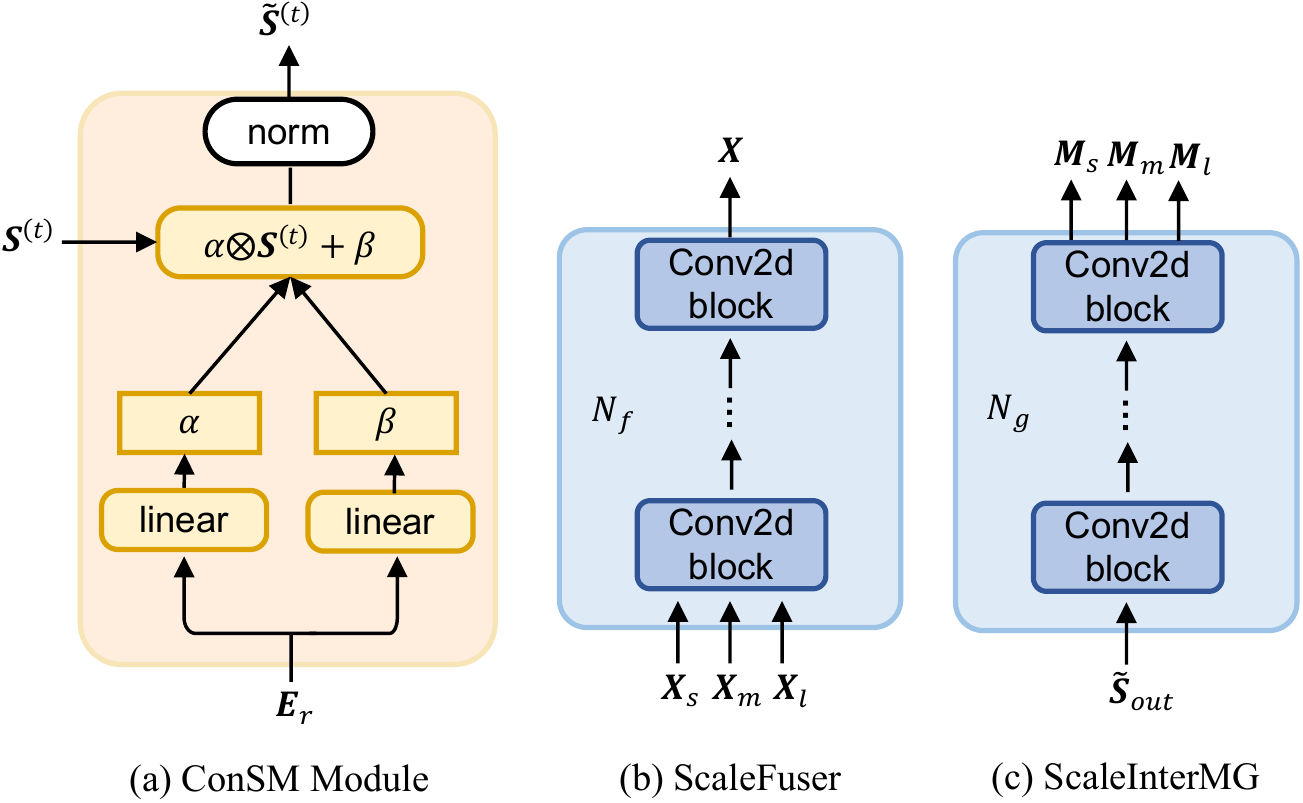}
    \vspace{-0.4cm}
	\caption{(a) The details of the \fusmodule{}, where ``$\otimes$" indicates the element-wise product.
	(b) The details of the \scaleencmodule{}.
	(c) The details of the \scaledecmodule{}.
	}

	\label{fig:parts}
	\vspace{-0.4cm}
\end{figure}

\subsection{Multi-Scale Fuser}
The notable successes of Conformer in source separation \cite{li2022ead, chen2021continuous} and speaker verification \cite{zhang2022mfa} evince that the effective fusion of temporal information of speech on multiple time scales \cite{toledano2018multi, chen2022fullsubnet+} enables neural networks to exploit richer acoustic features, which is conducive to the performance of the model.
However, the fusion of the multi-scale information in SpEx+ is quite rough.
It simply concatenates $\mathbf{S}_l, \mathbf{S}_m$ and $\mathbf{S}_s$ in the feature dimension, and fuses them using a 1-D convolution with kernel size of 1.
This approach fails to consider the local information of the feature dimension as well as the information in neighboring frames, and is not enough to sufficiently fuse multi-scale features.

As a consequence, we propose the more efficient multi-scale feature fusion module \scaleencmodule{} as an alternative to the original coarse method.
The structure of \scaleencmodule{} is shown in Figure \ref{fig:parts}(b), which consists of $N_f$ stacked Conv2d blocks, each of them contains 2-D convolution, an activation function ELU \cite{kim2020elastic}.
The module takes the multi-scale features $\mathbf{X}_{s}$, $\mathbf{X}_{m}$ and $\mathbf{X}_{l}$ as input, where $\mathbf{X}$ stands for $\mathbf{S}$ or $\mathbf{R}$.
We consider $\mathbf{X}_{s}$, $\mathbf{X}_{m}$ and $\mathbf{X}_{l}$ as different channels, and then interactively fuse the features of different channels through the 2-D convolution kernels in Conv2d blocks. Eventually the feature $\mathbf{X}$ that adequately combines information from different scales is obtained:
\begin{equation}
  \mathbf{X} = \mathcal{F}( [\mathbf{X}_{s}, \mathbf{X}_{m}, \mathbf{X}_{l}]).  
\end{equation}
where $\mathcal{F}(\cdot)$ represents the mapping function defined by the \scaleencmodule{}.
Compared with the fusion in SpEx+, \scaleencmodule{} takes the information of local feature dimensions and adjacent frames into account through 2-D convolution, which enables a more effective fusion of multi-scale features.
Additionally, for the consistency of two subnets' feature space, we share the parameters of \scaleencmodule{}s in the multi-scale fusion speech encoders.

\subsection{Conditional Speaker Modulation Module}
\label{sec:conSM}
It has been pointed out that the speaker embedding, which is concatenated with the frame-level speech features, cannot be sufficiently exploited by the stacked TCN blocks in the SpEx+ \cite{wang2021neural}.
Meanwhile, in text-to-speech (TTS) works \cite{min2021meta, wu2022adaspeech}, a small condition network called conditional layer normalization is implemented to modulate the hidden representations for synthesizing the target speaker's speech.
Inspired by this and accounting for the differences between TTS and speaker extraction tasks, we present the \fusmodule{} for addressing the above speaker embedding under-utilization problem based on conditional layer normalization.

Specifically, the conditional layer normalization \cite{wu2022adaspeech} adopts speaker-conditional affine transformation to modulate the hidden features that are performed layer normalization \cite{ba2016layer} at first.
In contrast to TTS, in speaker extraction, the speech features extracted from the mixture contain the target speaker's speech information, in which case we believe that applying layer normalization first may result in the loss of such information.
According to this, we tune the locations of layer normalization and speaker-conditional affine transformation, and apply it to the speaker extraction task, which is what we call \fusmodule{}.
As in Figure \ref{fig:parts}(a), the \fusmodule{} takes speaker embedding $\mathbf{E}_r$ and the frame-level speech feature $\mathbf{S}^{(t)}$ in speech extractor as the inputs, where $t=1,...,T$ denotes the frame indices.
We feed $\mathbf{E}_r$ to a linear layer with mapping function $\mathcal{D}(\cdot)$ and another linear layer with mapping function $\mathcal{R}(\cdot)$ to gain the adaptive scale vector $\alpha$ and bias vector $\beta$ respectively:
\begin{equation}
  \alpha = \mathcal{D}(\mathbf{E}_r), \quad  \beta = \mathcal{R}(\mathbf{E}_r).
\end{equation}
We perform an affine transformation of $\mathbf{S}^{(t)}$ with $\alpha$ and $\beta$, then conduct a layer normalization on the result of it:
\begin{equation}
  \widetilde{\mathbf{S}}^{(t)} = Norm(\alpha \otimes \mathbf{S}^{(t)} + \beta).
\end{equation}
where $Norm$ is the layer normalization and $\otimes$ represents the element-wise product.
Through this, depending on the condition of the given speaker embedding, we are able to scale up or down the speech features, negate them, and selectively set thresholds on them. 
Consequently, we can reinforce the information in the speech features that are relevant for extracting the target speaker in the \fusmodule{} module, instead of relying on the subsequent stacked TCNs to handle this, which resolves the speaker embedding under-utilization issue of the stacked TCNs in SpEx+.
The conditional affine transformation in \fusmodule{} may be relatively similar to FiLM \cite{perez2018film, rikhye2022closing, lei2022glow}, but as described above, the problem it seeks to solve is different from FiLM.
Furthermore, compared to FiLM, \fusmodule{} has an extra layer normalization operation after the affine transformation, which we argue is more appropriate for the speaker extraction task.

\subsection{Multi-Scale Interactive Mask Generator}
Previous works \cite{xu2020spex, ge2020spex+, wang2021neural} all apply three independent branches to generate three scales of receptive masks.
This leads to the issue that, when generating the mask of corresponding scale, each branch solely utilizes the information of one scale.
To this end, we introduce \scaledecmodule{}.
Through the interaction of different scales, it allows the model to make reasonable use of the valid information of other scales when generating the mask at a certain scale.

As shown in the Figure \ref{fig:parts}(c), the \scaledecmodule{} is composed of $N_g$ stacked Conv2d blocks, with each block containing a 2-D convolution, an ELU \cite{kim2020elastic} and a layer normalization.
The module takes $\widetilde{\mathbf{S}}_{out}$ as input, and then interactively fuses the features of different channels through the 2-D convolution kernel in the Conv2d block, ultimately treating the obtained features of three channels as multi-scale masks $\mathbf{M}_{s}$, $\mathbf{M}_{m}$ and $\mathbf{M}_{l}$:
\begin{equation}
  \mathbf{M}_{s}, \mathbf{M}_{m}, \mathbf{M}_{l} = \mathcal{H}(\widetilde{\mathbf{S}}_{out}).  
\end{equation}
where $\mathcal{H}(\cdot)$ is the mapping function that is intended to describe the \scaledecmodule{}.
By means of the above, \scaledecmodule{} can generate the mask at a certain scale with valid information from multiple scales, while reducing the number of parameters required by SpEx+ that employed three branches.

\section{Experiments}

\subsection{Datasets}
We conduct our experiments on the popular Libri2Mix dataset \cite{cosentino2020librimix}.
The \textit{train-100} subset, which is used for training, contains a total of 58 hours of utterances from 291 speakers.
The \textit{dev} subset and \textit{test} subset consist of 40 unseen speakers respectively, with an overall audio duration of 11 hours in each subset.
The \textit{dev} subset is served as the validation set during model training, while the \textit{test} subset is for the evaluation of the model's final performance.
For all speech audio, the sampling rate is 8 kHz.
Moreover, all of mixtures are in the `minimum' mode.

\subsection{Training Setup and Baselines}
We employ Adam optimizer with an initial learning rate of 1e-3. 
The learning rate decays by 0.5 once the performance on the validation set is not improved in 3 consecutive epochs.
The training of the model will be stopped when the best model is not found in the validation set after 8 consecutive epochs.
During the training procedure, both the mixed speech and the reference speech of target speaker are sliced into 3-second segments, while the full-length audio is applied at inference.

To testify the effectiveness of our improvement, the following models were compared.
Aiming at a fair comparison, the identical experimental setup is implemented for each model.
(1) \textbf{SpEx+:} Following \cite{ge2020spex+}, the convolutional filtering lengths of the speech encoder and decoder in SpEx+ are \{2.5, 10, 20\} ms respectively.
The number of ResNet blocks \cite{zeinali2019but} in ResNet speaker encoder is set to 3, and the dimension of speaker embedding is 256.
The hyperparameters are $M=4$ and $N=8$ for speaker-guided stacked TCNs in speech extractor.
The SpEx+ has a total of 11.78 M parameters.
(2) \textbf{\model{}:} 
There are 4 Conv2d blocks in \scaleencmodule{} with 2-D convolution channels of \{3, 32, 32, 1\} and kernal sizes of \{3, 3, 3, 3\}. 
The \scaledecmodule{} contains 4 Conv2d blocks having kernal sizes of \{3, 3, 3, 3\}, and their 2-D convolution channels are \{1, 32, 32, 3\}. 
The other configurations are the same as described above.
The number of parameters for \model{} is 10.77 M.

In our experiments, we mainly evaluate the model performance through three objective metrics, SI-SDR, PESQ, and ESTOI. 
Among them, SI-SDR is measured from the signal perspective, while PESQ and ESTOI are considered from the perceptual quality perspective, and the higher values of them are all positively correlated with better outcomes.


\begin{table}[!htbp]
    \begin{center}
    \caption{The performance in terms of SI-SDR [dB], PESQ [MOS] and ESTOI [\%] on the Libri2Mix test set.}
    \vspace{-0.2cm}
    \label{table:large total compare}
    \scalebox{0.9}{
    \begin{tabular}{c| c | c | c}
    \toprule
    Methods  &  SI-SDR & PESQ & ESTOI\\
    \hline

    Mixture    & 0.001 & 1.603 & 53.8 \\    
    TD-SpeakerBeam\cite{delcroix2020improving}   & 12.86 & 2.750 & - \\
    sDPCCN\cite{han2022dpccn}    & 11.65 & 2.738 & 78.9 \\
    TD-SpeakerBeam + PL$_2$+ PF$^{lin}$\cite{zhao2022target} & 13.88 & 2.860 & - \\
    \hline
    SpEx+\cite{ge2020spex+}   & 13.41 & 2.936 & 82.4 \\
    \model{}  & \textbf{14.61} & \textbf{3.195} & \textbf{84.9} \\
    \bottomrule
    \end{tabular}}
    \end{center}
\end{table}

\begin{table}[!htbp]
    \begin{center}
    \caption{Performance of SI-SDR and PESQ in section \ref{sec:overall_ablation_study} using the Libri2Mix test set.
    ``SF" and ``SIMG" mean \scaleencmodule{} and \scaledecmodule{} respectively.}
    \vspace{-0.3cm}
    \label{table:ablation}
    \scalebox{1.0}{
    \begin{tabular}{c|c|c|c|c}
    \toprule
    \multirow{2}{*}{\shortstack{Ablation\\Modules}}  & \multirow{2}{*}{ID} & \multirow{2}{*}{Settings} & \multirow{2}{*}{\shortstack{SI-SDR}} & \multirow{2}{*}{PESQ}\\
     & & & & \\
    \hline

       & \#1 & SpEx+ & 13.41 & 2.936 \\    

    \hline

    \multirow{2}{*}{\shortstack{Weight-share\\ \scaleencmodule{}s}}   
               & \multirow{2}{*}{\#2} & \multirow{2}{*}{\shortstack{\#1 + weight-\\share SFs}} & \multirow{2}{*}{14.05} & \multirow{2}{*}{3.075} \\
               &  &  &  & \\
    \hline
    \multirow{2}{*}{\scaledecmodule{}}   & \#3 & \#1 + SIMG & 13.72 & 3.059 \\    
               & \#4 & \#2 + SIMG & 14.32 & 3.153 \\
    \hline
    \multirow{2}{*}{\shortstack{\fusmodule{} \\ module}}   & \#5 & \#1 + \fusmodule{} & 13.69 & 2.969 \\    
               & \#6 & \#4 + \fusmodule{} & \textbf{14.61} & \textbf{3.195} \\           
    \bottomrule
    \end{tabular}}
    \end{center}
    \vspace{-0.6cm}
\end{table}

\begin{table}[!htbp]
    \begin{center}
    \caption{Performance of SI-SDR and PESQ in the investigation of weight-share \scaleencmodule{}s with the Libri2Mix test set.
    The ``$Enc_{Ext}$" and ``$Enc_{Spk}$" respectively denote the multi-scale fusion speech encoder in the subnet of speech extractor and speaker encoder.}
    \vspace{-0.3cm}
    \label{table:ablation scalefuser}
    \scalebox{1.0}{
    \begin{tabular}{l | c | c}
    \toprule
    Methods  &  SI-SDR & PESQ \\
    \hline
    SpEx+  & 13.41 & 2.936 \\
    \,\,+ \scaleencmodule{} in $Enc_{Ext}$ & 13.55 & 2.955 \\
    \,\,\,\,+ \scaleencmodule{} in $Enc_{Spk}$ & 13.70 & 3.031 \\
    \,\,\,\,\,\,+ shared weights  & \textbf{14.05} & \textbf{3.075} \\
    \bottomrule
    \end{tabular}}
    \end{center}
    \vspace{-0.9cm}
\end{table}

\subsection{Comparison with Baseline and State-of-the-art Methods}
\vspace{-0.1cm}
Table \ref{table:large total compare} shows the performance of different speaker extraction methods on the Libri2Mix test dataset.
In the last two rows of the Table, the performances of SpEx+ and the proposed \model{} are compared, and the results show that \model{} outperforms the SpEx+ in all evaluation metrics.
This indicates that our improvements have indeed improved the model's ability to extract the speech of target speakers.

We further compare the proposed \model{} to some other top-ranked methods
\cite{delcroix2020improving, han2022dpccn, zhao2022target} 
on the Libri2Mix dataset
in Table \ref{table:large total compare}.
Among them, the sDPCCN is a frequency-domain method with powerful capability \cite{han2022dpccn}.
While the TD-SpeakerBeam + PL$_2$ + PF$^{lin}$ has resolved the target confusion using the strategies of prototypical loss (PL) and post-filtering (PF) \cite{zhao2022target}, which is the previous best speech extraction system on the Libri2Mix dataset.
It can be concluded that comparing with these latest methods, \model{} achieves the state-of-the-art results on the Libri2Mix dataset. 

\subsection{Ablation Studies}
\vspace{-0.1cm}

In this section, taking SpEx+ as backbone, we perform a series of ablation studies about our improvements.

\vspace{-0.1cm}
\subsubsection{Overall Analysis on Proposed Modules}
\vspace{-0.1cm}
\label{sec:overall_ablation_study}
To start with, we analyze the contribution of our proposed modules to the model and the compatibility among them.
From Table \ref{table:ablation}, we can observe that appending weight-share \scaleencmodule{}s, \scaledecmodule{} and \fusmodule{} individually to SpEx+ shows certain effectiveness (\#2, \#3 and \#5).
It follows that the interfusion of multi-scale information in \scaleencmodule{} and \scaledecmodule{}, as well as using the \fusmodule{} to modulate, are beneficial to the model's performance.
Furthermore, all of the weight-share \scaleencmodule{}s, \scaledecmodule{} and \fusmodule{} do not conflict with each other (\#2, \#4 and \#6) and the combination of the three, the \model{}, achieves the best results (\#6).

\begin{table}[!htbp]
    \begin{center}
    \caption{The performance in terms of SI-SDR and PESQ in the investigation of \fusmodule{} module using the Libri2Mix test set.
    The ``Conditional LN" refers to the conditional layer normalization.}
    \vspace{-0.3cm}
    \label{table:ablation consm}
    \scalebox{1.0}{
    \begin{tabular}{l | c | c}
    \toprule
    Methods  &  SI-SDR & PESQ \\
    \hline
    SpEx+  & 13.41 & 2.936 \\
    SpEx+ with Conditional LN \cite{wu2022adaspeech} & 13.42 & 2.935 \\
    SpEx+ with FiLM \cite{rikhye2022closing} & 13.55 & 2.955 \\
    SpEx+ with \fusmodule{} module  & \textbf{13.69} & \textbf{2.969} \\
    \bottomrule
    \end{tabular}}
    \end{center}
    \vspace{-0.9cm}
\end{table}

\subsubsection{Investigation of Weight-share \scaleencmodule{}s}
\label{sec:ablation_study_sf}
We then further explore the role of the weight-share \scaleencmodule{}s.
As can be seen from Table \ref{table:ablation scalefuser}, the performance of the model has been improved after the additions of \scaleencmodule{} to the multi-scale fusion speech encoder in the subnet of speech extractor and speaker encoder.
This illustrates that our proposed \scaleencmodule{} indeed enables more effective fusion of multi-scale features than the original method in SpEx+.
Moreover, after sharing the weights of the \scaleencmodule{}s, the model performance is further boosted, which confirms effectiveness of our strategy to ensure the consistency of two subnets' feature space in the multi-scale feature fusion stage.

\subsubsection{Investigation of \fusmodule{} Module}
\label{sec:ablation_study_consm}
In order to investigate the effects of the \fusmodule{} module, in table \ref{table:ablation consm}, we compare it with the approaches mentioned in section \ref{sec:conSM}.
One can see that, after applying a dedicated module to exploit speaker embedding, the usage of either the FiLM or \fusmodule{} module brings a performance gain respectively, which indicates that this strategy does improve the SpEx+ with insufficient utilization of speaker embedding.
However, the addition of conditional layer normalization leads to almost no change in performance.
This is probably because that, as analyzed in subsection \ref{sec:conSM}, the normalization on the speech feature first operated in the conditional layer normalization introduces an information loss, which renders it inappropriate for the speaker extraction task.
In addition, the SpEx+ with \fusmodule{} module exceeds the SpEx+ with FiLM in terms of both the signal (SI-SDR) and perceptual quality (PESQ).
This also demonstrates that, compared to FiLM, our proposed \fusmodule{} is more suitable for speaker extraction.

\vspace{-0.25cm}
\section{Conclusions}
In this paper, we propose a framework with multi-scale interfusion and conditional speaker modulation named \model{}.
It adopts the weight-sharing \scaleencmodule{}s to effectively exploit the multi-scale information and ensure the consistency of the two subnets' feature space.
And then, the \scaledecmodule{} is presented to take the different scale information into account while generating masks.
Furthermore, we introduce \fusmodule{} module to fully blend the speaker embedding in the speech extractor.
Experimental results\footnote{Demo page: \href{https://rookiejunchen.github.io/MC-SpEx_demo}{https://rookiejunchen.github.io/MC-SpEx\_demo}} on the Libri2Mix dataset show that \model{} achieves a impressive performance and achieves state-of-the-art results for the speaker extraction task, which demonstrate the effectiveness of our improvements.


\textbf{Acknowledgement}: This work is supported by National Natural Science Foundation of China (62076144), Shenzhen Key Laboratory of next generation interactive media innovative technology (ZDSYS20210623092001004), Tencent AI Lab Rhino-Bird Focused Research Program (RBFR2022005) and Tsinghua University - Tencent Joint Laboratory.

\bibliographystyle{IEEEtran}
\bibliography{mybib}

\end{document}